\documentclass[11pt]{article}

\usepackage{array}
\usepackage{graphicx}
\usepackage{amsmath}
\usepackage{amssymb}
\usepackage{hhline}

\title{\bf Convergence conditions for iterative methods seeking multi-component
solitary waves with prescribed quadratic conserved quantities}

\author{ T.I. Lakoba\footnote{lakobati@cems.uvm.edu, \ 1 (802) 656-2610} 
 \vspace{0.5cm} \\
  Department of Mathematics and Statistics, 16 Colchester Ave., \\
 University of Vermont, Burlington, VT 05401, USA}

\newcommand{\noi}{\noindent}
\newcommand{\D}{\Delta}
\newcommand{\be}{\begin{equation}}
\newcommand{\ee}{\end{equation}}
\newcommand{\bsube}{\begin{subequations}}
\newcommand{\esube}{\end{subequations}}
\newcommand{\ba}{\begin{array}}
\newcommand{\ea}{\end{array}}
\newcommand{\To}{\rightarrow}
\newcommand{\vecx}{{\bf x}}
\newcommand{\vecy}{{\bf y}}

\newcommand{\vecb}{{\bf b}}
\newcommand{\vecr}{{\bf r}}
\newcommand{\vecd}{{\bf d}}
\newcommand{\vecu}{{\bf u}}

\newcommand{\Lnot}{L^{(0)}}
\newcommand{\Lnnot}{L^{(00)}}
\newcommand{\Lcal}{{\mathcal L}}

\newcommand{\Knnot}{K^{(00)}}
\newcommand{\Kcal}{{\mathcal K}}

\newcommand{\vecQ}{\vec{Q}}
\newcommand{\vecmu}{\vec{\mu}}
\newcommand{\vecB}{\vec{B}}
\newcommand{\vecH}{\vec{H}}
\newcommand{\vecphi}{\vec{\varphi}}
\newcommand{\Ucal}{{\mathcal U}}
\newcommand{\Vcal}{{\mathcal V}}

\newcommand{\bea}{\begin{eqnarray}}
\newcommand{\eea}{\end{eqnarray}}
\newcommand{\so}{\Rightarrow}
\newcommand{\dst}{\displaystyle}

\newcommand{\tu}{\tilde{u}}
\newcommand{\tv}{\tilde{v}}

\newcommand{\vecN}{{\bf N}}

\newcommand{\eq}{Eq.}
\newcommand{\eqs}{Eqs.}

\renewcommand{\theequation}{\thesection.\arabic{equation}}

\begin{document}
\baselineskip 18 pt

\maketitle

\vspace*{2cm}

\begin{center}
 {\bf Abstract}
\end{center}

We obtain local (i.e., linearized) convergence conditions for iterative methods that seek
solitary waves with prescribed values of
quadratic conserved quantities of multi-component Hamiltonian nonlinear wave equations.
These conditions extend the ones found for 
single-component solitary waves in [J. Yang and T.I. Lakoba, Stud. Appl. Math. {\bf 120}, 265--292 (2008)].
We also show that, and why, these convergence conditions coincide with dynamical stability conditions
for ground-state solitary waves.

\vskip 1.1 cm

\noi
{\bf Keywords}: \ Coupled nonlinear wave equations,
Solitary waves, Iterative methods.

\bigskip

\noi
{\bf PACS}: \ 03.75.Lm, 05.45.Yv, 42.65.Tg, 47.35.Fg.

\newpage


{\bf
Solitary wave solutions of most nonlinear wave equations can be found only numerically.
Recently, J. Yang and the present author obtained \cite{ITEM} 
conditions under which an iterative numerical method can converge to
stationary solitary waves of single-component Hamiltonian nonlinear wave equations. When this method,
in what follows referred to as the imaginary-time evolution method (ITEM), converges, 
it provides one with a numerical approximation of a solitary waves with a prescribed value
of a quadratic conserved quantity usually referred to either as {\em power} or the number of particles.
However, many phenomena are described not by a single equation but by systems of coupled equations.
Therefore, it is of interest to obtain conditions under which a multi-component counterpart
of the ITEM would be guaranteed to converge to, i.e., find, a multi-component solitary wave. We obtain such
a condition in this work. Moreover, generalizing an observation made in \cite{ITEM}, we show that
the multi-component ITEM converges only to those ground states of nonlinear wave equations 
which are dynamically stable, and explain why this is the case.
 }

\section{Introduction and background}

For the single-equation case considered in Ref. \cite{ITEM}, the power of a solitary wave is
\be
P=\int u^2(\vecx)\,d\vecx\,,
\label{e1_01}
\ee
where $u$ is the real-valued field of the solitary wave and $\vecx$ is the spatial coordinate. 
(Here and below, if the limits of the integration are not indicated, the
integration is assumed to be over the entire spatial domain.) For example, if the time-dependent wave 
$U(\vecx,t)$ satisfies a Nonlinear Schr\"odinger-type (NLS-type) equation
\be
iU_t+\nabla^2 U + G(|U|^2,\vecx)\,U=0\,, \qquad U(|\vecx|\To \infty) \To 0\,, 
\label{e1_02}
\ee
where $\nabla^2$ is the Laplacian, then upon the substitution \
$U(\vecx,t)=e^{i\mu t}u(\vecx)$, where $u$ is real, \eq~\eqref{e1_02} reduces to
\bsube
\be
\Lnnot u - \mu u \equiv \Lnot u = 0, 
\label{e1_03a}
\ee
where
\be
\Lnnot u \equiv \nabla^2 u + G(u^2,\vecx)\,u\,.
\label{e1_03b}
\ee
\label{e1_03}
\esube
The parameter $\mu$ is referred to as the propagation constant of the solitary wave.
A straightforward calculation shows that the power $P=\int u^2 d\vecx = \int |U|^2 d\vecx$ 
is conserved by the evolution equation \eqref{e1_02}. 
Thus, $u$ can be parametrized either by $P$ or by $\mu$, so that one can write $P \equiv P(\mu)$.

The method analyzed in \cite{ITEM} finds the solution $u$ with a prescribed value of $P$
by iterations:
\bsube
\be
\mu_n= \frac{ \langle N^{-1}u_n, \Lnnot u_n \rangle}{ \langle N^{-1}u_n, u_n \rangle } \,,
\label{e1_04a}
\ee
\be
\hat{u}_{n+1} - u_n = N^{-1} \left( \Lnnot u_n - \mu_n u_n \right) \D\tau\,,
\label{e1_04b}
\ee
\be
u_{n+1}=\hat{u}_{n+1} \sqrt{ \frac{P}{\langle \hat{u}_{n+1}, \hat{u}_{n+1} \rangle } } \,.
\label{e1_04c}
\ee
\label{e1_04}
\esube
where $\D\tau>0$ is an auxiliary parameter, and a positive definite operator $N$ 
can be conveniently chosen in the form mimicking the linear constant-coefficient part of 
$\Lnot$ in \eqref{e1_03a}:
\be
N=c-\nabla^2, \qquad c>0\,.
\label{e1_05}
\ee
The purposes of $ \D\tau$ and $N^{-1}$ in \eqref{e1_04b} will be clarified shortly.
The inner product is defined as
\be
\langle f(\vecx), g(\vecx) \rangle  \,\equiv\, \int (f(\vecx))^T g(\vecx) \, d\vecx\,.
\label{e1_06}
\ee

Methods similar to \eqref{e1_04} had also been considered for finding solitary waves in the past
(see, e.g., \cite{Chiofalo00}--\cite{ShchesnovichC04}). However, it was in \cite{ITEM}
where the convergence conditions of the specific version, \eqref{e1_04}, of the ITEM were
found in terms of the properties of the linearized operator of \eq~\eqref{e1_03a}. Namely, 
\eqs~\eqref{e1_04} are linearized by a substitution
\be
u_n=u+\tu_n, \qquad |\tu_n| \ll u,
\label{e1_07}
\ee
which results in \cite{ITEM}
\be
\tu_{n+1} - \tu_n = N^{-1} \Lcal \tu_n \D\tau, \qquad 
\Lcal \tu_n \, \equiv \,  L \tu_n - 
   \frac{ \langle N^{-1}u, L\tu_n \rangle}{ \langle N^{-1} u, u \rangle }\,u \,.
\label{e1_08}
\ee
Here $L$ is the linearized operator of $\Lnot$ in \eq~\eqref{e1_03a}. For example, 
for $\Lnnot$ given by \eqref{e1_03b},
\be
L=\nabla^2 - \mu + G(u^2, \vecx) + 2u^2 G_{u^2}(u^2, \vecx)\,.
\label{e1_09}
\ee
The ITEM \eqref{e1_04} converges if the eigenvalues of the right-hand side of its linearization
\eqref{e1_08} are located between $-2$ and $0$. The first of these conditions implies that
\be
\D\tau < -2/\Lambda_{\min},
\label{e1_10}
\ee
where $\Lambda_{\min}$ is the most negative eigenvalue of $N^{-1} \Lcal$.
(In practice, the maximum $\D\tau_{\max}$ can be easily found by just a few trials, and
then the value $\D\tau$ leading to an optimal convergence rate
is usually somewhat smaller than $\D\tau_{\max}$; see Fig.~3 and \eq~(47) in \cite{ITEM}.)
The second condition 
%
%
can be shown \cite{ITEM} to be related to the
properties of the original equation \eqref{e1_03a} and its linearized operator $L$ 
as follows \cite{footnote1}.
First, let us denote 
\be
p(L)\,\equiv \,\mbox{the number of positive eigenvalues of $L$ counting their multiplicity}
\label{e1_11}
\ee
(and similarly for any other operator or matrix).
Next, assume that we are considering the generic case whereby
the null space of $L$ does not contain any functions other than those of
the modes $\partial u/\partial x^{(k)}$ which  are associated with translational invariance
of the solitary wave along $x^{(k)}$. 
Then, under condition \eqref{e1_10}, the ITEM \eqref{e1_04} converges provided that
either
\bsube
\be
p(L)=0 \quad {\rm and} \quad \partial P/\partial \mu \neq 0\,,
\label{e1_12a}
\ee
or
\be
p(L)=1 \quad {\rm and} \quad \partial P/\partial \mu > 0\,.
\label{e1_12b}
\ee
\label{e1_12}
\esube
In all the other cases algorithm \eqref{e1_04} diverges \cite{footnote2}.
Remarkably, as pointed out in \cite{ITEM}, these convergence conditions are the same as
the stability conditions of nodeless solitary waves in the NLS-type
evolution equation \eqref{e1_02} \cite{VK}.
In other words, the ITEM \eqref{e1_04} converges only to those nodeless solitary waves
of \eqref{e1_02} that are dynamically linearly stable. (Let us note, in passing, that
iterative methods that are guaranteed to converge to {\em any} solitary wave, stable or
unstable, were first proposed in \cite{GarciaRipollPG01} and later developed in \cite{SOM}.)

The purpose of using operator $N^{-1}$ in \eqref{e1_04b} is to considerably reduce the
magnitude of the most negative eigenvalue of operator $N^{-1}\Lcal$ compared to that of $L$
\cite{GarciaRipollPG01, ITEM, SOM}. This is analogous to preconditioning a poorly conditioned
matrix $A$ when solving a linear equation $A\vecy=\vecb$ by an iterative method;
this can considerably improve the convergence rate of the iterations
(see, e.g., \cite{TrefethenBau_NLA}, Lecture 40). 
In regards to implementing $N$, note that
it is a differential operator with constant coefficients and hence has a simple
representation in the Fourier space. Therefore, $N^{-1}\Lnnot u_n$ and $N^{-1}u_n$
are easily computed using the direct and inverse Fast Fourier Transforms,
which are available as built-in commands in all major computing software.

Note that the propagation constant $\mu$ in the ITEM \eqref{e1_04} is not prescribed but
computed iteratively. 
Numerical methods for finding solutions of Eq.~\eqref{e1_03} with a specified value of $\mu$
rather than with a specified value of power \eqref{e1_01} have also been considered in quite
a few studies (see, e.g., references in \cite{CGM}), and we will not consider them in this
work.

In this paper, we derive convergence conditions of a generalization of the ITEM \eqref{e1_04}
for multi-component solitary waves. Such a generalization was proposed in \cite{CGM} (but
see also Section 4.3 in \cite{SOM}), 
and its algorithm is presented in Section 2. Note that in \cite{CGM}, we also
proposed another method that finds solitary waves with the same conserved quantities
as the ITEM but converges faster; moreover, it converges {\em much} faster than the ITEM when the
latter converges slowly. This method is a modified form of the well-known Conjugate Gradient method 
(CGM; see, e.g., \cite{TrefethenBau_NLA}, Lecture 38), and its algorithm is given in Appendix
for the reader's convenience.
In \cite{CGM} we showed that this modified CGM has the same convergence conditions as the
ITEM. Therefore, these convergence conditions, which generalize conditions \eqref{e1_12},
apply to both the ITEM and modified CGM. The derivation of these conditions is the main
result of this paper and is presented in Section 3. This derivation is based on the idea
of Ref. \cite{Pelinovsky2000}, where it was used to establish {\em stability} conditions for
a certain class of multi-component solitary waves. In fact, our convergence conditions
of the ITEM and modified CGM turn out to be the same as
the stability conditions derived in \cite{Pelinovsky2000}.
This relation between the two sets of conditions generalizes a similar fact pointed out
in \cite{ITEM} for single-component equations, and in Section 5 we explain under what
circumstances such a coincidence of the convergence and stability conditions occurs. 
In Section 4 we provide a geometrical argument that facilitates intuitive interpretation 
of the convergence conditions derived in Section 3 for
the special case where the solitary wave has two quadratic conserved quantities.
In Section 6 we summarize the results of this work.
Let us emphasize that numerical examples involving the multi-component ITEM and CGM
are not a subject of this analytical study; the interested reader can find such examples in Ref. \cite{CGM}.


\section{ITEM algorithm for multi-component solitary waves}
\setcounter{equation}{0}

Let us begin with an example that will motivate introduction of some new notations.
The following system describes pulse
evolution in a two-core nonlinear directional fiber coupler where each core supports two 
orthogonal polarizations of light \cite{DCDP}:
\be
\ba{l}
\dst 
iU^{(1)}_t+U^{(1)}_{xx}+ \left( |U^{(1)}|^2+\kappa |U^{(2)}|^2 \right) U^{(1)} + U^{(3)}= 0 
\vspace{0.2cm} \\
\dst
iU^{(2)}_t+U^{(2)}_{xx}+ \left( |U^{(2)}|^2+\kappa |U^{(1)}|^2 \right) U^{(2)} + U^{(4)}= 0
\vspace{0.2cm} \\
\dst 
iU^{(3)}_t+U^{(3)}_{xx}+ \left( |U^{(3)}|^2+\kappa |U^{(4)}|^2 \right) U^{(3)} + U^{(1)}= 0 
\vspace{0.2cm} \\
\dst 
iU^{(4)}_t+U^{(4)}_{xx}+ \left( |U^{(4)}|^2+\kappa |U^{(3)}|^2 \right) U^{(4)} + U^{(2)}= 0 
\ea
\label{e2_01}
\ee
Here $(U^{(1)},U^{(2)})$ and $(U^{(3)},U^{(4)})$ are the pairs of orthogonal polarizations in
the two cores. The quadratic quantities conserved by these equations and generalizing \eqref{e1_01} are:
\be
\vecQ = \left( \ba{c} P^{(1)}+P^{(3)} \\ P^{(2)}+P^{(4)} \ea \right) \equiv
 \left( \ba{cccc} 1 & 0 & 1 & 0 \\ 0 & 1 & 0 & 1 \ea \right) \
 \left( \ba{c} P^{(1)} \\ P^{(2)} \\ P^{(3)} \\ P^{(4)} \ea \right)\,,
\label{e2_02}
\ee
where \ $P^{(k)}=\langle (U^{(k)})^*,U^{(k)} \rangle\,$. \ 
Upon the substitution 
\be
\left( \ba{c} U^{(1)}(x,t) \\ U^{(2)}(x,t) \\ U^{(3)}(x,t) \\ U^{(4)}(x,t) \ea \right)  = 
\left( \ba{cc} u^{(1)}(x) & 0 \\ 0 & u^{(2)}(x)  \\ u^{(3)}(x) & 0 \\ 0 & u^{(4)}(x) \ea \right)
\left( \ba{c} \dst e^{i\mu^{(1)}t} \\ \dst e^{i\mu^{(2)}t} \ea \right)\,,
\label{e2_04}
\ee
where $u^{(k)}$ can (for the purpose of this example) 
be chosen to be real-valued, system \eqref{e2_01} reduces to:
\be
\left( \ba{l} u^{(1)}_{xx}+ \left( (u^{(1)})^2+\kappa (u^{(2)})^2 \right) u^{(1)} + u^{(3)} 
              \vspace{0.1cm} \\
              u^{(2)}_{xx}+ \left( (u^{(2)})^2+\kappa (u^{(1)})^2 \right) u^{(2)} + u^{(4)} 
              \vspace{0.1cm} \\
              u^{(3)}_{xx}+ \left( (u^{(3)})^2+\kappa (u^{(4)})^2 \right) u^{(3)} + u^{(1)} 
              \vspace{0.1cm} \\
              u^{(4)}_{xx}+ \left( (u^{(4)})^2+\kappa (u^{(3)})^2 \right) u^{(4)} + u^{(2)} 
              \ea \right)  \; - \; 
   \left( \ba{cc} u^{(1)} & 0 \\ 0 & u^{(2)} \\ u^{(3)} & 0 \\ 0 & u^{(4)} \ea \right)\,\vecmu
    \;=\;  \left( \ba{c} 0 \\ 0 \\ 0 \\ 0 \ea \right)   \,, 
   \label{e2_05}
   \ee
where $\vecmu= \left( \mu^{(1)},\, \mu^{(2)} \right)^T$. Then the powers of the
individual components are 
\be
P^{(k)}=\langle u^{(k)},u^{(k)} \rangle\,.
\label{e2_03}
\ee

We now generalize this example to an $S$-component system possessing an $s$-component 
vector of conserved quantities $\vecQ$, so that the $k$th component of $\vecQ$ is:
\be
Q^{(k)}= \sum_{l=1}^S q^{(kl)} P^{(l)}, 
\qquad k=1,\ldots\,, \,s \le S\,, \quad l=1,\ldots\,,\, S,
\label{e2_06}
\ee
where 
%
%
the solitary wave is 
$\vecu= \big( u^{(1)},\ldots\,, u^{(S)} \big)^T$.
As illustrated in the above example with $S=4$ and $s=2$, 
the number of conserved quantities can be less than the number of
the components of the solitary wave: $s\le S$. 
Other examples where the situation $s<S$ takes place include:
a system of NLS-type equations coupled coherently via phase-sensitive {\em nonlinear}
terms (as opposed to linear ones as in \eqref{e2_01}); 
the well-known system of three waves interacting via 
quadratic nonlinearity \cite{3w_76, 3w_02}; and any system of coupled carrier-wave (also 
known as long-wave, or Korteweg--de Vries-type (KdV-type)) equations, as we will explain
at the end of Section 5.
To emphasize the possibility of having $s < S$, we use a different
vector notation for $\vecQ$ than for ${\bf u}$.
Now, the matrix $\big( q^{(kl)} \big)$ in \eqref{e2_06} is assumed to be in reduced echelon form
(see any textbook on undergraduate Linear Algebra) and, in addition, its columns are arranged
so that
\be
q^{(kl)} \,=\, \left\{ \ba{ll} 0, & l<k \\ 1, & l=k\,. \ea \right. 
\label{e2_07}
\ee
In \eqref{e2_02}, matrix $\big( q^{(kl)} \big)$ is the $2\times 4$ matrix. 

The multi-component generalization of \eqs~\eqref{e1_03a} and \eqref{e1_04a} is
\bsube
\be
{\bf \Lnot u} \,\equiv \, 
{\bf \Lnnot}{\bf u} - \Ucal \, \langle {\bf N}^{-1} \Ucal,\, \Ucal \rangle^{-1}\,
  \langle {\bf N}^{-1} \Ucal, \, {\bf \Lnnot} {\bf u} \rangle \,=\, {\bf 0}\,,
\label{e2_08a}
\ee
\be
\Ucal \equiv \frac{\delta \vecQ}{\delta {\bf u}}\,.
\label{e2_08b}
\ee
\label{e2_08}
\esube
For example, in \eqref{e2_05}, ${\bf \Lnnot u}$ is the first term (the $4\times 1$ vector), and 
$\Ucal$ is the first factor of the second term (the $4\times 2$ matrix) on the left-hand side.
The $S\times S$ matrix ${\bf N}$ is a self-adjoint positive definite operator. For optimal
preconditioning, its differential part should mimic the highest derivative in the linear
part of ${\bf \Lnnot}$. For example, for the ${\bf \Lnnot}$ in \eqref{e2_05}, ${\bf N}$ can
be chosen as a diagonal matrix with its diagonal entries of the form \eqref{e1_05}.

The multi-component version of the ITEM \eqref{e1_04} is:
\bsube
\be
\hat{\bf u}_{n+1} - {\bf u}_n  \,=\, {\bf N}^{-1} \left( {\bf \Lnnot u}_n - 
 \Ucal_n \, \langle {\bf N}^{-1} \Ucal_n,\, \Ucal_n \rangle^{-1}\,
  \langle {\bf N}^{-1} \Ucal_n, \,  {\bf \Lnnot} {\bf u}_n \rangle  \right) \D\tau\,,
\label{e2_09a}
\ee
\be
u^{(k)}_{n+1} = \hat{u}^{(k)}_{n+1} 
\sqrt{ \frac{ Q^{(k)}- \sum_{l=k+1}^S q^{(kl)} \hat{P}^{(l)}_{n+1} }{ \hat{P}^{(k)}_{n+1} } }\;, 
\qquad    k=1,\ldots\,,\, s\le S\,,
\label{e2_09b}
\ee
\label{e2_09}
\esube
where
$$
\hat{P}^{(k)}_{n+1} \equiv \langle \hat{u}^{(k)}_{n+1},\, \hat{u}^{(k)}_{n+1} \rangle\,,
\qquad  k=1,\ldots\,,\,S\,.
$$
Note that, by \eqref{e2_07}, the numerator of the fraction under the square root equals $P^{(k)}$.
Let us emphasize that if $s<S$, then \eqref{e2_09b} specifies only that the $s$ components of
$\vecQ$ have their prescribed values but does not impose any other conditions on the
powers, $P^{(k)}$, of the {\em individual} components of the solitary wave.

As we noted in Section 1, in \cite{CGM} we proposed a modified CGM that converges 
under the same conditions that we will establish for the ITEM \eqref{e2_09}, but faster.
Moreover, it converges {\em much} faster when the ITEM converges slowly. 
Its algorithm, however, is somewhat less transparent than \eqref{e2_09}, and therefore
we state it in Appendix. Let us reiterate: The convergence analysis that we will
present in Section 3 applies both to the ITEM and the modified CGM. We advocate using
the latter method when the slow convergence of the ITEM justifies spending a little
extra effort on programming the algorithm of the CGM.

As in the single-component case, we perform convergence analysis of the ITEM \eqref{e2_09}
using linearization analogous to \eqref{e1_07}. 
A tedious but straightforward calculation shows that 
the linearized operator of the right-hand side of \eq~\eqref{e2_09a} is:
\be
{\bf N}^{-1} {\mathcal L}{\bf \tu}_n  \equiv {\bf N}^{-1} \left( {\bf L \tu}_n - 
\Ucal \, \langle {\bf N}^{-1} \Ucal,\, \Ucal \rangle^{-1}\,
  \langle {\bf N}^{-1}\Ucal, \, {\bf L} {\bf \tu}_n \rangle\, \right) \,.
\label{e2_10}
\ee
Here ${\bf L}$ is the linearized operator of ${\bf \Lnot}$ in \eqref{e2_08a}
obtained when the last term in that equation is replaced by \ $\Ucal\,\vecmu$; compare
with \eqref{e2_05}. (Although $\vecmu$ is not prescribed but instead is iteratively
computed within the method, its exact value can still be used in the convergence analysis.)
For Hamiltonian wave equations, ${\bf L}$ is self-adjoint.
Next, the conservation of $\vecQ$ implies the orthogonality relation
\be
\langle {\Ucal,\, \bf \tu}_n \rangle = \vec{0}\,.
\label{e2_11}
\ee
Taking the inner product between $\Ucal$ and \eqref{e2_09a}, one can show that 
\eq~\eqref{e2_09b} does not change the linearization of \eqref{e2_09a};
the role of \eqref{e2_09b} is to guarantee that the $s$ components of vector $\vecQ$ equal
their prescribed values {\em exactly} rather than in the linear approximation.
Thus, the operator in \eqref{e2_10} is the linearized
operator of the multi-component ITEM. Operator $\Lcal$ is easily verified \cite{ITEM}
to be  self-adjoint on the space of
functions satisfying the orthogonality relation \eqref{e2_11}. 
However, ${\bf N}^{-1} \Lcal$ is not self-adjoint.
To cast the linearized ITEM \eqref{e2_09} into a form involving only self-adjoint
operators, which is more convenient to analyze than \eqref{e2_10}, 
we use the following change of variables:
\be
{\bf \tv}_n = {\bf N}^{1/2}{\bf \tu}_n, \qquad \Vcal={\bf N}^{-1/2}\Ucal, \qquad 
\Kcal = {\bf N}^{-1/2}\Lcal {\bf N}^{-1/2},  \qquad  
{\bf K} =  {\bf N}^{-1/2}{\bf L}{\bf N}^{-1/2}\,.
\label{e2_12}
\ee
Then the linearized ITEM \eqref{e2_10} and the orthogonality relation \eqref{e2_11}
become:
\be
{\bf \tv}_{n+1} - {\bf \tv}_n = \Kcal {\bf \tv}_n \D\tau, \qquad 
\Kcal {\bf \tv}_n \,\equiv\, {\bf K \tv}_n - 
 \Vcal \, \langle \Vcal,\, \Vcal \rangle^{-1}\,
 \langle \Vcal, \, {\bf K} {\bf \tv}_n \rangle\,,
 \label{e2_13}
\ee
\be
\langle {\Vcal,\, \bf \tv}_n \rangle = \vec{0}\,.
\label{e2_14}
\ee

In what follows we will analyze the transformed form \eqref{e2_13} of the linearized ITEM,
because it involves operator $\Kcal$ that is self-adjoint on the space of functions 
satisfying the orthogonality relation \eqref{e2_14}. Therefore,
the evolution of the iteration error ${\bf \tv}_n$ is completely determined by the 
eigenvalues of $\Kcal$. For convergence of the ITEM, these eigenvalues must lie between
$-2/\D\tau$ and $0$. The first of these conditions is achieved by adjusting $\D\tau$,
whereas the second condition is analyzed in the next Section, where we will establish its 
connection to the number of positive eigenvalues of ${\bf L}$. It should be pointed out
that by Sylvester's law of inertia (see, e.g., \cite{HornJohnson_book}), the numbers
of positive and zero eigenvalues of ${\bf K}$ and ${\bf L}$ are the same. Therefore, we
will refer everywhere to those eigenvalues of ${\bf L}$, whereas in the analysis of
\eqref{e2_13} it is the eigenvalues of ${\bf K}$ that are involved directly.

Finally, a note is in order about the effect of zero eigenvalues of ${\bf L}$.
As in \cite{ITEM}, we assume the generic situation whereby the null space of ${\bf L}$ 
does not contain any functions other than those of
the modes $\partial {\bf u}/\partial x^{(k)}$ which  are associated with translational 
invariance of the solitary wave along coordinate $x^{(k)}$. As was shown in \cite{ITEM}
and \cite{CGM} for the single-component ITEM and CGM, such modes lead only to a slight shift 
of the solitary wave along the respective coordinates, but do not otherwise affect
convergence of the iterative method. The same proofs carry over directly to the 
multi-component case. Thus, in what follows we will focus on nonzero eigenvalues of ${\bf L}$.

\setcounter{equation}{0}
\section{Stability criterion for multi-component iterative methods}

This Section contains the main result of this work, Eqs. \eqref{e3_01}, which are derived
using a combination of analyses of Refs. \cite{ITEM} and \cite{Pelinovsky2000}; see \cite{footnote3}.
Namely, we will show that the operator $\Kcal$ is
negative definite on the space of functions satisfying \eqref{e2_14} provided that
the Jacobian matrix
\bsube
\be
\frac{\partial \vecQ}{\partial \vecmu} \equiv 
  \frac{\partial (Q^{(1)},\ldots\,, Q^{(s)} )}{\partial (\mu^{(1)},\ldots\,, \mu^{(s)} )}
 \qquad \mbox{is nonsingular}
\label{e3_01a}
\ee
and that
\be
p({\bf L}) = p \left( \frac{\partial \vecQ}{\partial \vecmu} \right)\,,
\label{e3_01b}
\ee
\label{e3_01}
\esube
where the notation $p$ is defined in \eqref{e1_11}. These conditions 
generalize conditions 
\eqref{e1_12} for the multi-component ITEM \eqref{e2_09} and modified CGM \eqref{A1_01}. 
As explained at the end of Section 2, under
these conditions the ITEM can be guaranteed to converge by choosing $\D\tau$ to be sufficiently 
small (in practice, $\D\tau=O(1)$ \cite{ITEM,SOM}). The modified CGM is guaranteed to converge provided that 
\eqref{e3_01} hold.

Let $\Psi$ be an eigenfunction of $\Kcal$ and $\Phi$ be an eigenfunction of ${\bf K}$:
\be
\Kcal \Psi = \Lambda \Psi, \qquad {\bf K}\Phi = \lambda \Phi\,.
\label{e3_02}
\ee
Taking the inner product of $\Vcal$ with the first equation in \eqref{e3_02}
and using the definition of $\Kcal$, one sees that eigenfunctions $\Psi$ with $\Lambda\neq 0$
satisfy the orthogonality relation \eqref{e2_14}. However, the eigenfunction $\Psi$ 
with $\Lambda=0$ does {\em not}, in general, satisfy that relation, as we will see later on.
Let us now expand $\Psi$ and $\Vcal$ over the set of $\Phi$'s:
\be
\ba{l}
\dst
\Psi = \sum_m a_m \Phi_m(\vecx) + \int_{\rm cont} a(\lambda) \Phi(\lambda,\vecx) \,d\lambda\,, \\
\dst
\Vcal = \sum_m \Phi_m(\vecx) \vecB_m^T + \int_{\rm cont} \Phi(\lambda,\vecx)\vecB^T(\lambda) \,d\lambda\,.
\ea
\label{e3_03}
\ee
Here the two terms in each expansion correspond to the contributions of the discrete
and continuous spectra of ${\bf K}$, and $a$'s are scalars and 
$\vecB$'s are $s\times 1$ vectors: 
\be
\vecB_m = \langle \Vcal, \Phi_m \rangle, \qquad \vecB(\lambda) = \langle \Vcal, \Phi(\lambda) \rangle\,.
\label{e3_04}
\ee
Here and below we do not indicate the dependence of $\Phi$ and $\Psi$ on $\vecx$ since it
is always implied. Let us also denote an $s\times 1$ vector 
\be
\vecH = \langle \Vcal,\, \Vcal \rangle^{-1}\,
 \langle \Vcal, \, {\bf K} \Psi(\Lambda) \rangle\,.
 \label{e3_05}
\ee
 From the \eqref{e3_02}, \eqref{e3_03}, \eqref{e3_05} one finds:
\be
a_m = \vecB^T_m \vecH /(\lambda_m-\Lambda), \qquad
a(\lambda) = \vecB^T(\lambda) \vecH /(\lambda-\Lambda).
\label{e3_06}
\ee
Substitution of these equations into the orthogonality relation \eqref{e2_14}, 
which is to be satisfied by $\Psi(\Lambda)$ for $\Lambda\neq 0$, yields:
\be
R\, \vecH =\vec{0}, \qquad R(\Lambda) \equiv 
\sum_m \frac{\vecB_m \vecB^T_m}{\lambda_m-\Lambda} + 
\int_{\rm cont} \frac{\vecB(\lambda) \vecB^T(\lambda)}{\lambda-\Lambda} \,d\lambda\,.
\label{e3_07}
\ee

Before continuing, we need to point out one important feature of the eigenvalue problem
for $\Psi(\Lambda)$, which can be restated as
\be
{\bf K}\Psi - \Lambda\Psi = \Vcal \vecH\,.
\label{e3_add01}
\ee
Namely, the vector $\vecH$ in it is {\em arbitrary}, and, therefore, by specifying
different $\vecH$'s one obtains different $\Psi$'s for a given $\Lambda$. 
To verify this, one only needs
to substitute ${\bf K}\Psi$ from \eqref{e3_add01} into \eqref{e3_05}.
This observation about $\vecH$ being arbitrary allows one to reformulate the problem
$R(\Lambda)\vecH=\vec{0}$ as follows:
\be
\mbox{\em Analyze under what conditions matrix $R(\Lambda)$ can
be singular for $\Lambda>0$.} 
\label{e3_add02}
\ee
If we find that $R(\Lambda)$ can be singular only for $\Lambda<0$, 
this would imply that $\Kcal$ is negative definite (modulo the remark made at the end of
Section 2) and hence the multi-component ITEM and modified CGM would converge.
Again, note that in arriving at formulation \eqref{e3_add02}, we have relied on the
arbitrariness of $\vecH$. Indeed, if $\vecH$ had not been arbitrary but instead determined
by $\Psi(\Lambda)$, then the first equation in \eqref{e3_07} would not have been equivalent
to \eqref{e3_add02}, since even though $R(\Lambda)$ could have been singular, the particular
$\vecH$ might not have necesarily been its eigenvector.

To address question \eqref{e3_add02}, 
we study the eigenvalue problem for the $s\times s$ real and symmetric matrix $R$:
\be
R \vecphi = \gamma(\Lambda) \vecphi\,.
\label{e3_08}
\ee
Its eigenvalues can be found from the Rayleigh quotient:
\be
\gamma(\Lambda) = \vecphi^T R \vecphi \,/\, \vecphi^T \vecphi\,.
\label{e3_09}
\ee
 From \eqref{e3_09}, \eqref{e3_07}, \eqref{e3_04} and the completeness of the set of $\Phi$'s one has:
\be 
\gamma(\Lambda\To +\infty) = -\langle \Vcal\vecphi,\Vcal\vecphi \rangle \,/\,
( \vecphi^T \vecphi\,  \Lambda)  \,\To \, -0\,.
 \label{e3_10}
 \ee
Here we have used the fact that rank of $\Vcal$ is $s$, since $\Vcal$ is constructed
from $s$ independent components of $\vecQ$; hence $\Vcal\vecphi\neq \vec{0}$. 
Similarly, one verifies that
all the eigenvalues of $R$ satisfy
\be
d\gamma/d\Lambda < 0 \quad {\rm when} \quad \Lambda \neq \lambda_m.
\label{e3_11}
\ee
As $\Lambda\To \lambda_m$, the matrix $(\lambda_m-\Lambda)R \To \vecB_m \vecB^T_m$.
The latter is a rank-one matrix, and hence only one of its eigenvalues is nonzero.
Therefore, at $\Lambda=\lambda_m$, at most (see below) one eigenvalue of $R$ has a simple
pole singularity and the other eigenvalues are continuous functions of $\Lambda$.

\begin{figure}[t!]
\hspace*{-0.1cm}
\mbox{ 
\begin{minipage}{6cm}
\rotatebox{0}{\resizebox{6cm}{7.7cm}{\includegraphics[0in,0.5in]
 [8in,10.5in]{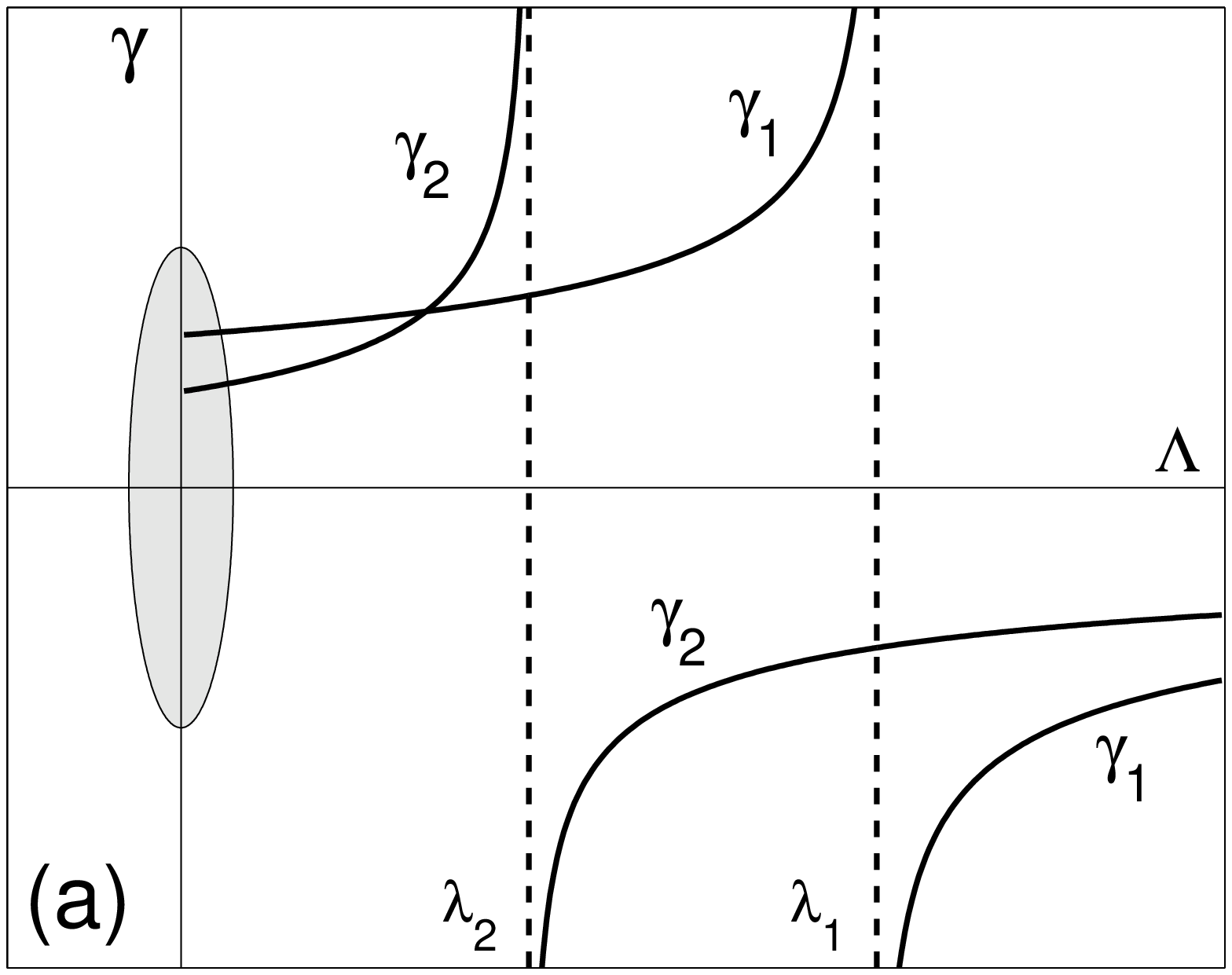}}}
\end{minipage}
\hspace{-0.1cm}
\begin{minipage}{6cm}
\rotatebox{0}{\resizebox{6cm}{7.7cm}{\includegraphics[0in,0.5in]
 [8in,10.5in]{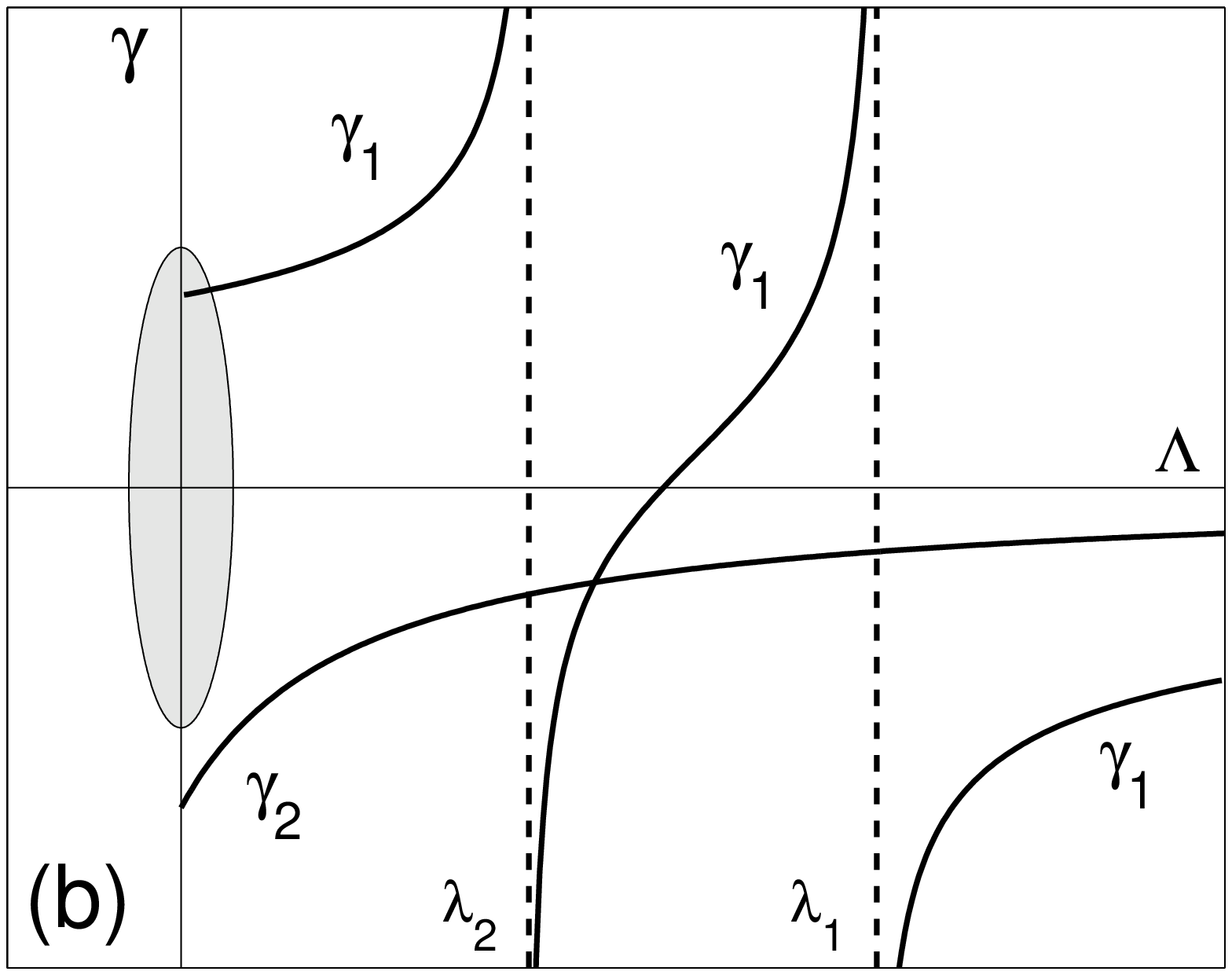}}}
\end{minipage}
 }
 
\vspace{-2.9cm}

\mbox{ 
\begin{minipage}{6cm}
\rotatebox{0}{\resizebox{6cm}{7.7cm}{\includegraphics[0in,0.5in]
 [8in,10.5in]{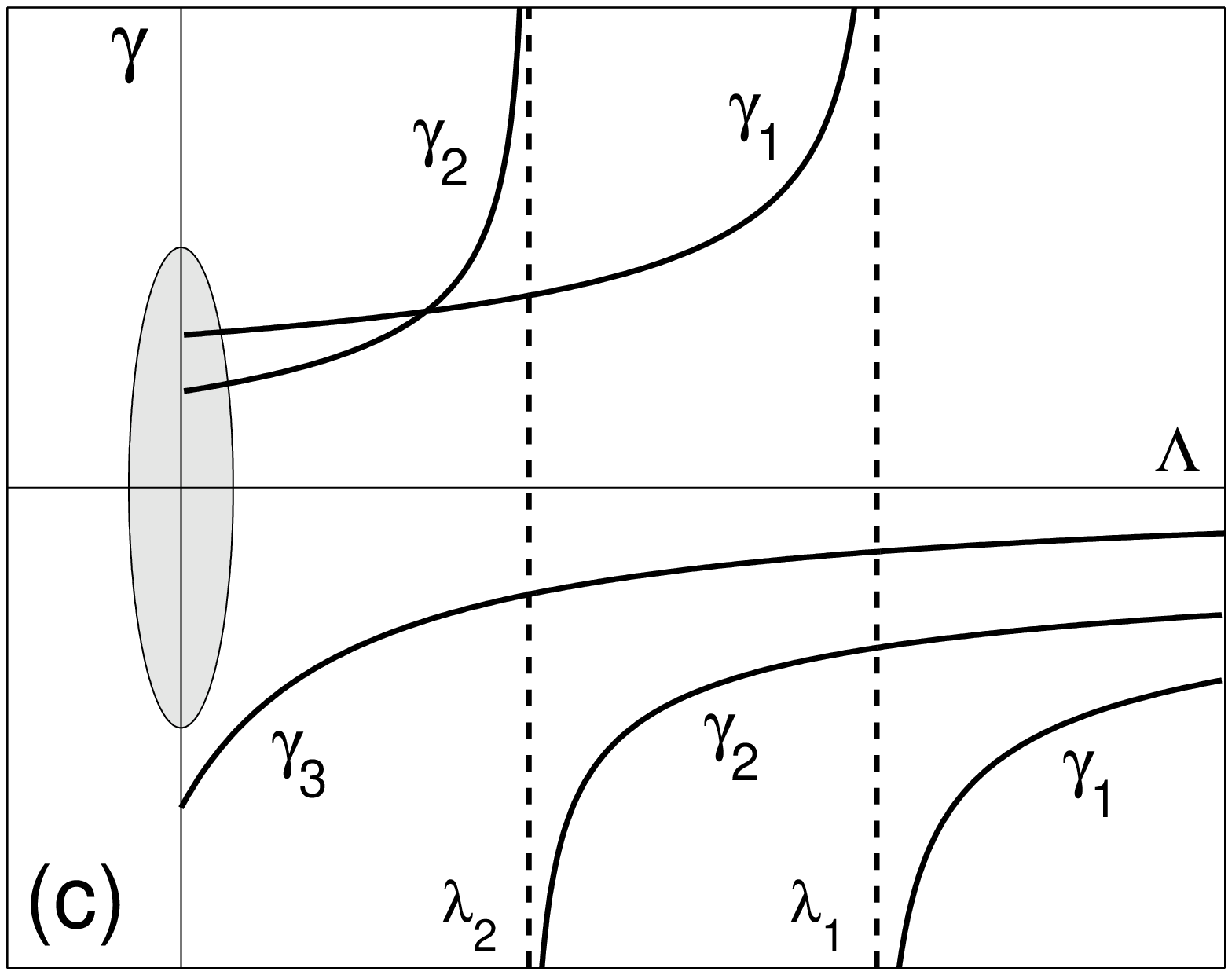}}}
\end{minipage}
\hspace{-0.1cm}
\begin{minipage}{6cm}
\rotatebox{0}{\resizebox{6cm}{7.7cm}{\includegraphics[0in,0.5in]
 [8in,10.5in]{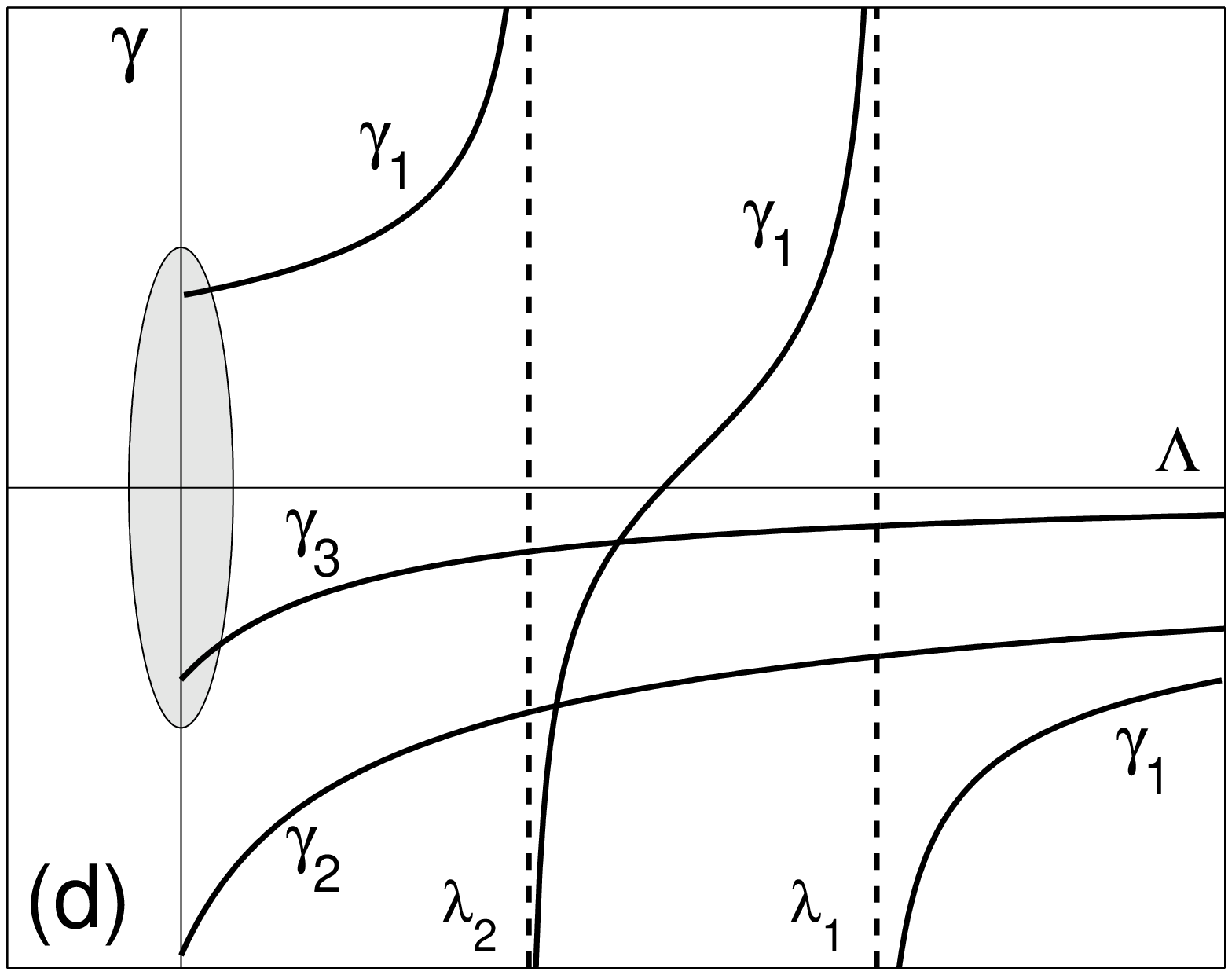}}}
\end{minipage}
 }
 
\vspace{-2.9cm}

\mbox{ 
\begin{minipage}{6cm}
\rotatebox{0}{\resizebox{6cm}{7.7cm}{\includegraphics[0in,0.5in]
 [8in,10.5in]{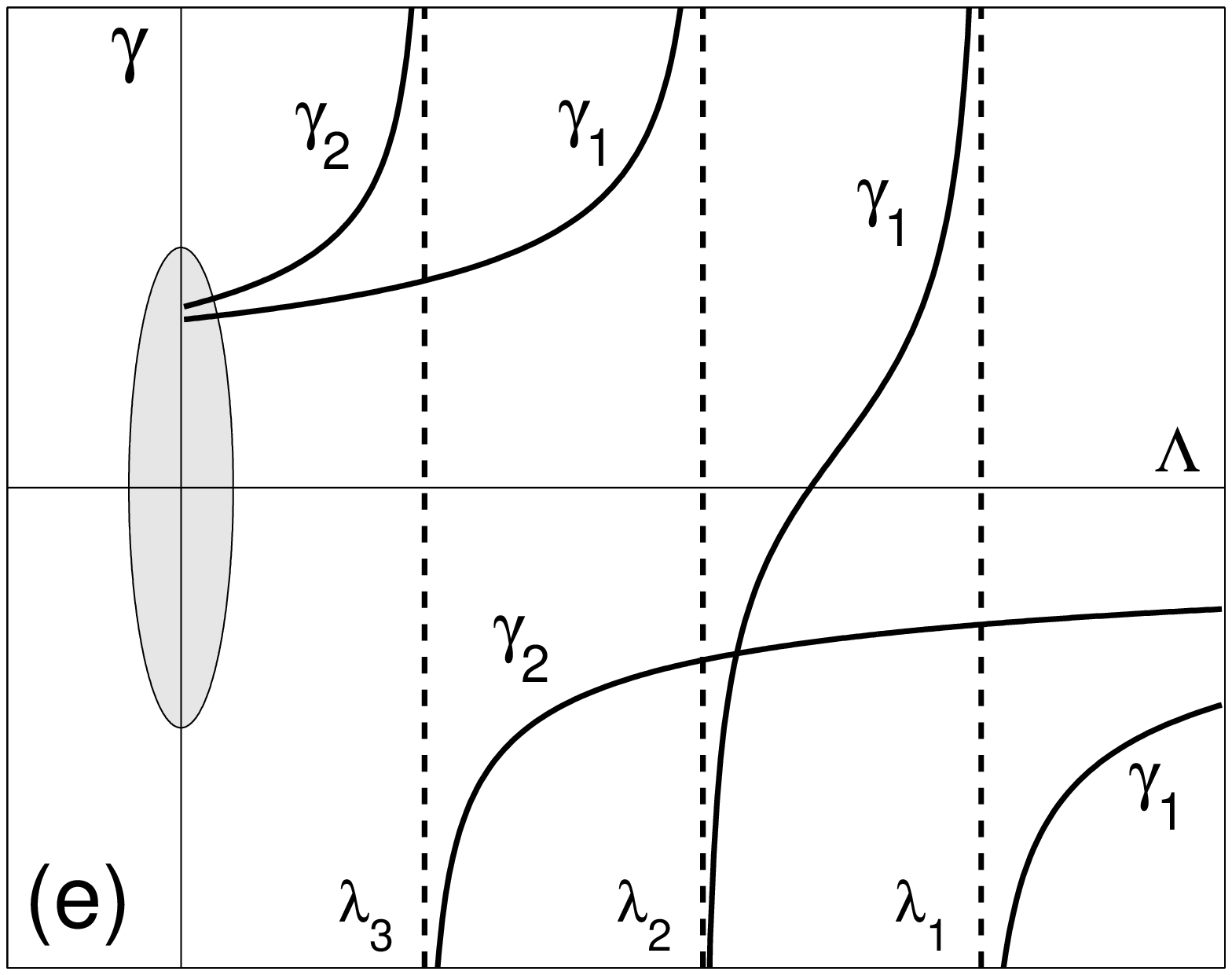}}}
\end{minipage}
\hspace{-0.1cm}
\begin{minipage}{6cm}
\rotatebox{0}{\resizebox{6cm}{7.7cm}{\includegraphics[0in,0.5in]
 [8in,10.5in]{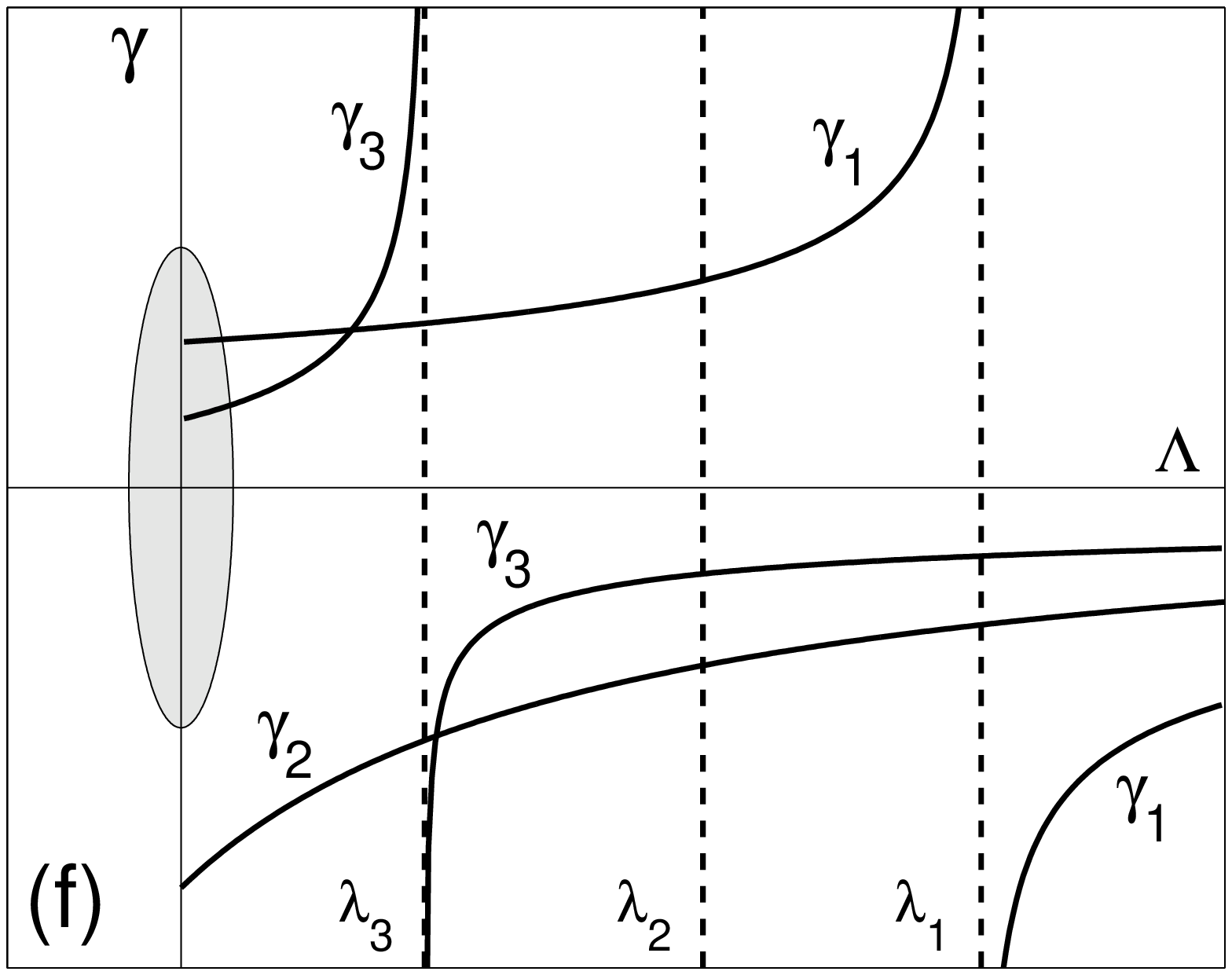}}}
\end{minipage}
 }
 
\vspace{-2.9cm}

\mbox{ 
\begin{minipage}{6cm}
\rotatebox{0}{\resizebox{6cm}{7.7cm}{\includegraphics[0in,0.5in]
 [8in,10.5in]{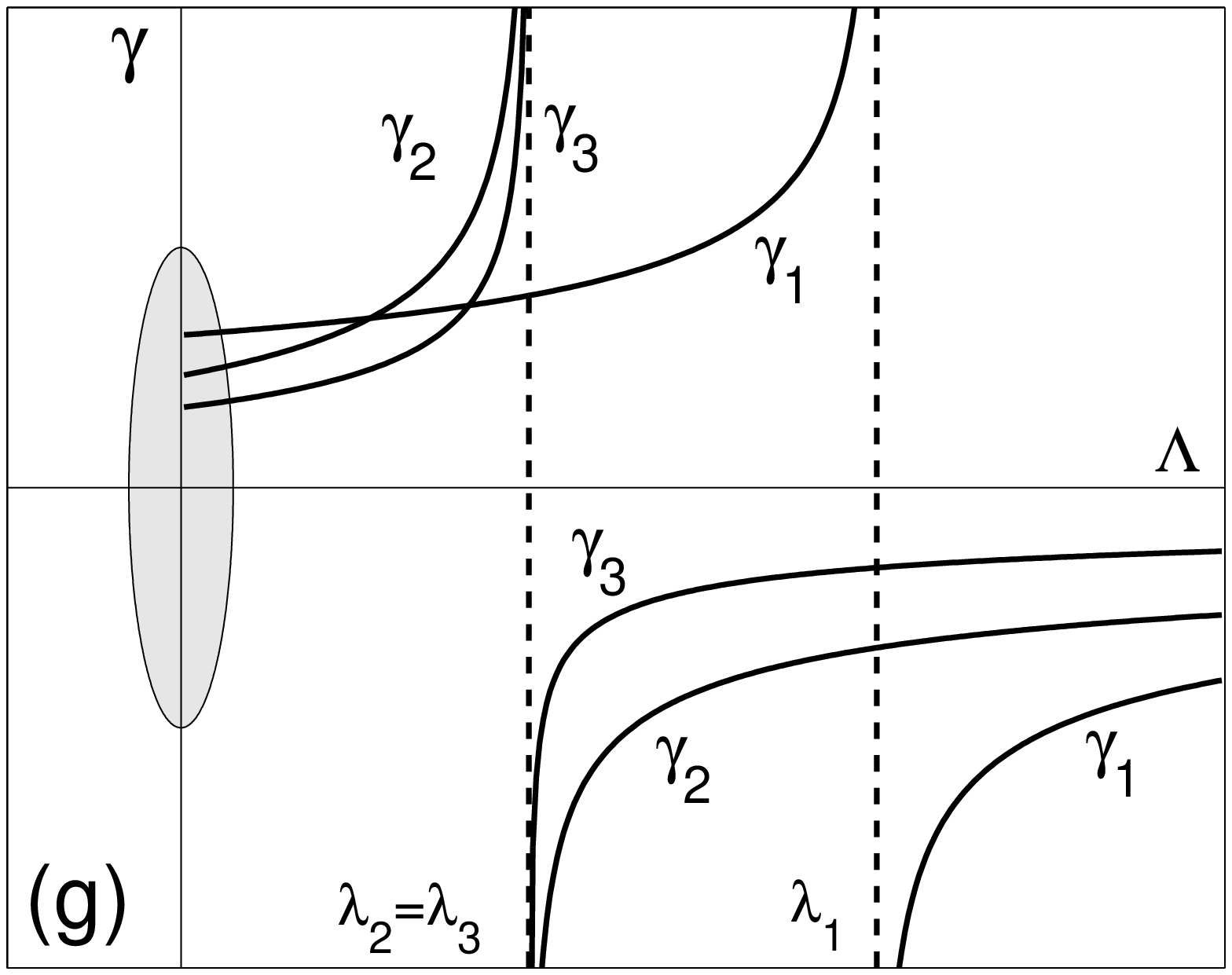}}}
\end{minipage}
\hspace{-0.1cm}
\begin{minipage}{6cm}
\rotatebox{0}{\resizebox{6cm}{7.7cm}{\includegraphics[0in,0.5in]
 [8in,10.5in]{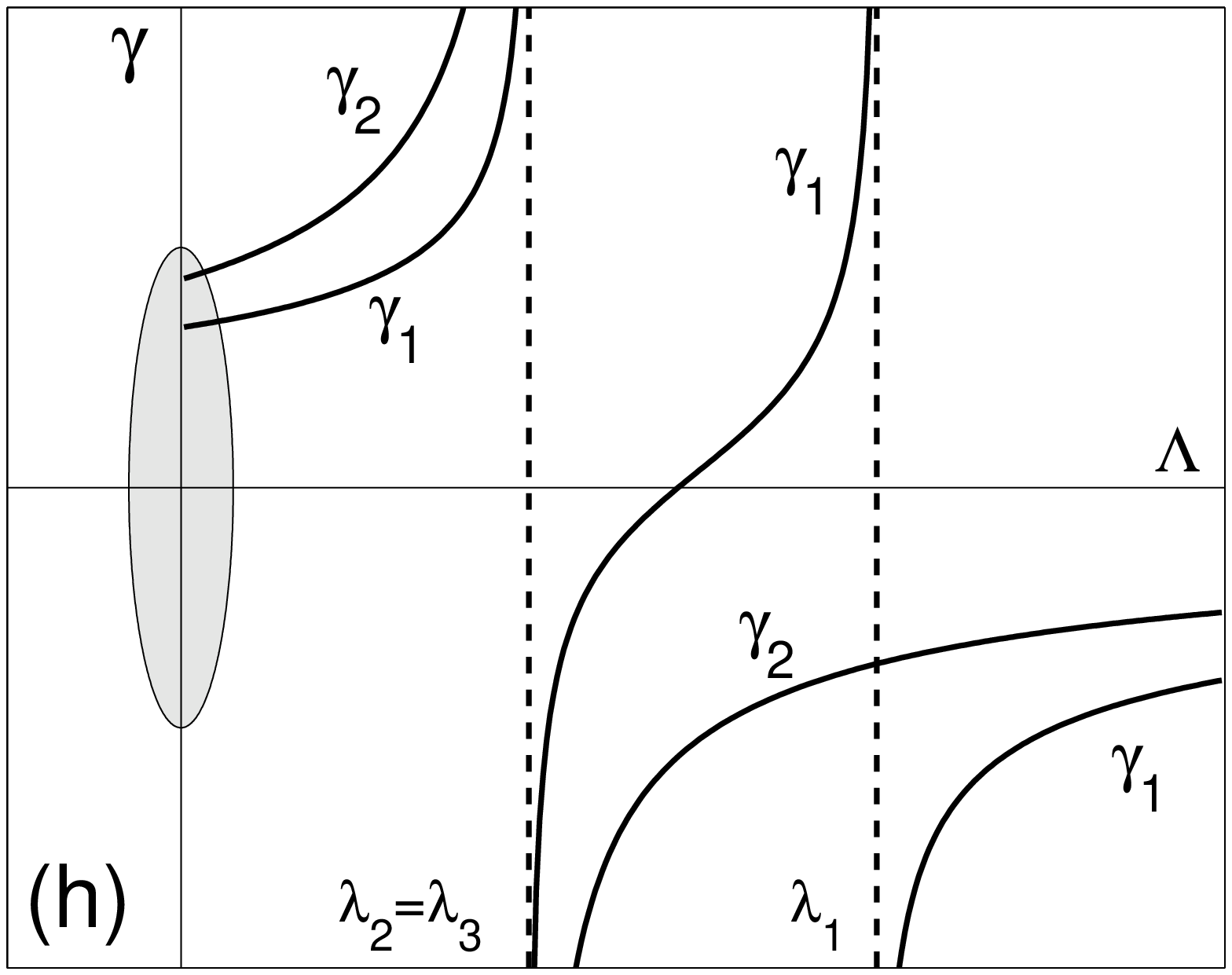}}}
\end{minipage}
 }
\vspace{-1.6cm}
\caption{Various cases showing the eigenvalues of matrix $R(\Lambda)$,
as explained in the text. The shaded oval along the vertical axis indicates
that the curves which are depicted as crossing that axis at positive values
can actually cross it anywhere (i.e., also at negative values). \ (a) $s=p({\bf K})=2$; \ 
(b) $s=p({\bf K})=2$; \ (c) $s=3,\;p({\bf K})=2$; \  (d) $s=3,\;p({\bf K})=2$; \  
(e) $s=2,\;p({\bf K})=3$; \  (f) $s=p({\bf K})=3$; \  
(g) $s=p({\bf K})=3$; \ (h) $s=2,\;p({\bf K})=3$. 
}
\label{fig_1}
\end{figure}

The facts stated in the previous paragraph allow one to specify when 
$R$ can be singular (i.e., one of $\gamma(\Lambda)=0$) for $\Lambda>0$. 
We will do so for generic cases first and at the end will consider the missed special cases.
One should consider three possibilities:
\be 
\mbox{(i) \quad $s=p({\bf K})$, \qquad (ii) \quad $s > p({\bf K})$, \qquad 
(iii) \quad $s<p({\bf K})$.}
\label{e3_12}
\ee
Qualitatively different situations for possibilities (i) and (ii) are examplified by
Fig.~\ref{fig_1}(a--d). From panels (a,c) in this Figure one can see that $\gamma=0$ does {\em not}
occur for $\Lambda>0$ provided that \ $p({\bf K}) = p(R(0))$. 
 From Fig.~\ref{fig_1}(b,d) it also follows that if \ $p({\bf K}) > p(R(0))$, then
there is always a $\gamma=0$ for some $\Lambda>0$. 
By inspection, one can
convince oneself that these statements are true is the general case (i.e., for any
$s$ and $p({\bf K})$) for possibilities (i) and (ii).
One can also see that the situation where \ 
$p({\bf K}) < p(R(0))$ \ cannot occur. Indeed, by \eqref{e3_10}, all 
$\gamma(\Lambda\To+\infty) < 0$, and they can become positive only at $\lambda_m$'s.
Hence the number of positive eigenvalues of $R(0)$ cannot exceed the number of 
positive $\lambda_m$'s, which is $p({\bf K})$.
Finally, in regards to possibility (iii), one can easily see (Fig.~\ref{fig_1}(e))
that there should always
be a $\gamma=0$ for some $\Lambda>0$. To summarize, $R$ does not become singular for
positive $\Lambda$ only when \ $p({\bf K}) = p(R(0))$.

Thus, to arrive at conditions \eqref{e3_01}, we need to relate $R(0)$ with 
$\partial\vecQ/\partial \vecmu$. First, in analogy to the first equation in \eqref{e3_07},
\be
R(0)\vecH = \langle \Vcal, \Psi(0) \rangle\,.
\label{e3_13}
\ee
As we noted after \eq~\eqref{e3_02}, the right-hand side of \eqref{e3_13} does not,
in general, vanish. We will now find $\Psi(0)$. From the definition of $\Kcal$,
\be
{\bf K}\Psi(0) = \Vcal \vecH\,.
\label{e3_14}
\ee
On the other hand, consider \eq~\eqref{e2_08a} written for the transformed operator
${\bf \Knnot v} \equiv {\bf N}^{-1/2} {\bf \Lnnot u}$ and note that the last
term in that equation is $\Vcal \vecmu$ (see also \eqref{e2_05}).
Differentiation of this equation with respect to $\mu^{(k)}$ yields:
\bsube
\be
{\bf K} \frac{\partial \, {\bf v}}{\partial \mu^{(k)} } = \Vcal {\bf e}^{(k)}\,, \qquad k=1,\ldots\,,s,
\label{e3_15a}
\ee
where ${\bf v}={\bf N}^{1/2}{\bf u}$ and
${\bf e}^{(k)}$ is the $s\times 1$ vector whose $k$th entry is 1 and the other entries are zero.
Combining \eqs~\eqref{e3_15a} for all $k$ yields
\be
{\bf K} \frac{\partial {\bf v} }{\partial \vecmu} = \Vcal \,, 
\label{e3_15b}
\ee
\label{e3_15}
\esube
where $\partial {\bf v} / \partial \vecmu$ is an $S\times s$ matrix. Now, comparison of
\eqref{e3_14} and \eqref{e3_15b} shows that
\be
\Psi(0)= \partial {\bf v} / \partial \vecmu + \sum_l g^{(l)}\,\partial {\bf v} / \partial x^{(l)},
\label{e3_16}
\ee
where $g^{(l)}$ are arbitrary constants, and we have used our assumption that the null space
of ${\bf L}$ (and hence of ${\bf K}$) can consist only of translational-invariance eigenmodes.
All such eigenmodes are orthogonal to $\Vcal$, as can be seen by considering their inner products with
\eqref{e3_15b}. Then the substitution of \eqref{e3_16} into \eqref{e3_13} and recalling that
 \ $\Vcal=\delta \vecQ/\delta {\bf v}$ \ (see \eqref{e2_08b} and \eqref{e2_12}) yields \ 
$R(0)\vecH = \partial\vecQ/\partial\vecmu \; \vecH$. Given the arbitrariness of $\vecH$
(see the text after \eqref{e3_add01}), this implies that 
\be
R(0) = \partial\vecQ/\partial\vecmu\,.
\label{e3_17}
\ee
This fact along with the summary sentence found before \eq~\eqref{e3_13} entails condition
\eqref{e3_01b} in the generic case.

Let us now consider special cases that we have glossed over. First, suppose one of the terms
in the discrete sum in \eqref{e3_07} with a $\lambda_m>0$ is a zero matrix. 
This can occur only when $\vecB=\vec{0}$.
Then by the first equation in \eqref{e3_04}, the corresponding eigenfunction $\Phi_m$ of 
${\bf K}$ satisfies the orthogonality condition \eqref{e2_14} and, by \eqref{e3_add01}
and \eqref{e3_05}, is also an eigenfunction of $\Kcal$ with the eigenvalue $\Lambda=\lambda_m>0$.
Thus, even though all the eigenvalues $\gamma$ of $R(\lambda)$ are continuous at $\Lambda=\lambda_m$
and do not change their signs (see Fig.~\ref{fig_1}(f), where $\lambda_m=\lambda_2$), 
operator $\Kcal$ still has a
positive eigenvalue $\Lambda=\lambda_m$. 
Note that since fewer than $p({\bf K})$ of the eigenvalues $\gamma$
change sign as $\Lambda$ decreases from $+\infty$ to $0$, then $p(R(0)) < p({\bf K})$ and
hence by \eqref{e3_17}, this special cases falls under the generic condition \eqref{e3_01b}.

Second, suppose that two positive eigenvalues of the self-adjoint operator ${\bf K}$ are the same. 
Since the corresponding eigenfunctions are linearly independent, so are the eigenvectors of
$R$ whose eigenvalues will have a pole singularity at $\Lambda=\lambda_m$. Figures \ref{fig_1}(g,h)
illustrate two qualitatively different situations that can occur in this case. As one can see,
condition \eqref{e3_01b} still determines whether any of the $\gamma(\Lambda)$'s can vanish
for $\Lambda>0$. 

Third, suppose $\partial\vecQ/\partial\vecmu = R(0)$ is singular. This means that there is
an eigenfunction $\Psi(0)$ of $\Kcal$ that satisfies the orthogonality condition \eqref{e2_14}.
As we point our below, this may prevent the ITEM and
modified CGM from converging. Hence we impose condition \eqref{e3_01a}. Thus, we have
established both conditions \eqref{e3_01} as being necessary and sufficient to guarantee that
the iterative methods \eqref{e2_09} (with \eqref{e1_10} being satisfied)
and \eqref{A1_01} will converge for any initial guess ${\bf u}_0$
that is sufficiently close to the solitary wave being sought.
 
Let us now discuss how these methods may behave if either of these conditions is violated.
If \eqref{e3_01b} does not hold, then the ITEM is guaranteed to diverge for a generic initial
condition, since the iteration error ${\bf \tu}_n$ will contain a component of the eigenmode
that will increase by a factor $(1+\Lambda\D\tau)>1$ at each iteration. On the other hand,
if \eqref{e3_01a} is violated, then the iteration error will contain an eigenmode with
$\Lambda=0$, which will remain the same at each iteration. Hence the ITEM will settle near
$({\bf u}+\mbox{small constant}\cdot \Psi(0))$, and the norm of the iteration error will not be
able to reach an arbitrarily low prescribed error. 
Thus, the linearized convergence analysis predicts that the method will ``stall" at a higher
error, but will not diverge. (Taking into account terms nonlinear in $\tu_n$ 
(see \eqref{e1_07}) may yield the information of whether the method actually converges or
diverges. However, such a nonlinear analysis is of limited practical use since, even if the
ITEM is found to eventually converge, it would do so very slowly in this case.)

As for the modified CGM, it is {\em not} bound to diverge 
if $\Kcal$ is not negative definite. However, it {\em may} do so when either of conditions
\eqref{e3_01} is violated. The mechanism of this divergence would be the vanishing of the
denominator in \eqs~(\ref{A1_01}b,e) of the algorithm. Such a divergence
may perhaps be avoided by choosing a different (but still generic) initial guess ${\bf u}_0$.

Finally, let us present an example where one can predict convergence of the ITEM and modified 
CGM without computing the eigenvalues of ${\bf L}$ and $\partial\vecQ/\partial\vecmu$.
This example is a straightforward extension of Corollary 1 in \cite{ITEM} to the 
multi-component case. Consider a system of incoherently coupled NLS-type equations,
generalizing \eqref{e1_02}:
\be
iU^{(k)}_t + \nabla^2 U^{(k)} + G^{(k)}\!\left(|U^{(1)}|^2,\ldots\,, |U^{(S)}|^2,\,\vecx \right) 
 \, U^{(k)} \,=\, 0.
\label{e3_18}
\ee
Its solitary wave is sought in the form $U^{(k)}(\vecx,t)=u^{(k)}(\vecx)\exp(i\mu^{(k)}t)$
with $u^{(k)}$ being real. 
Note that in this case, $s=S$, $\vecQ=\big( P^{(1)}, \ldots\,,P^{(S)}\big)$, and also
\bsube
\be
{\bf L} = {\bf \Lnot} + {\mathcal G}, \qquad
({\mathcal G})^{(kl)}=2u^{(k)}u^{(l)} \;\partial G^{(k)} / \partial \big( u^{(l)} \big)^2\,.
\label{e3_19a}
\ee
In this case the operator ${\bf \Lnot}$ is diagonal with entries
\be
\big( {\bf \Lnot} \big)^{(k)} = -\mu^{(k)} + \nabla^2  + G^{(k)}\,.
\label{e3_19b}
\ee
\label{e3_19}
\esube

\vspace{-0.5cm}

Suppose that at least one of the components of ${\bf u}$ has at least one node and 
${\mathcal G}$ is positive (semi)definite. Then $p({\bf L})>S$, and hence the iterative
methods \eqref{e2_09} and \eqref{A1_01} diverge. 

On the other hand, suppose that all components of ${\bf u}$ are nodeless and
${\mathcal G}$ is negative (semi)definite. Then $p({\bf L})=0$, and hence the iterative
methods converge. The proof of both statements repeats that of Corollary 1 in \cite{ITEM}
and hence is not given here.

\setcounter{equation}{0}
\section{Geometric interpretation of $p(\partial\vecQ/\partial\vecmu)$ when $s=2$}


In the case of a single equation, the ITEM's convergence condition $\partial P/\partial \mu > 0 $ in 
\eqref{e1_12b} has a simple geometric interpretation: the curve $P(\mu)$ must have a positive slope. 
(In other words, solitary waves for which $p(L)=1$ and curve $P(\mu)$ has a negative slope cannot
be obtained by the ITEM; however, they may be obtained by other iterative methods \cite{ITEM}.)
The convergence condition in the multi-component case, \eq~\eqref{e3_01b}, is not as straightforward
to visualize. In this section we will give a geometric interpretation of 
this condition for the case $s=2$, i.e. when the solitary wave has two quadratic
conserved quantities $Q^{(1)}$ and $Q^{(2)}$. (Note that the solitary wave in this case can have
more than two components: examples include the three-wave system \cite{3w_76, 3w_02} and 
\eqs~\eqref{e2_01}.)
While in the $s=1$ case the Jacobian $\partial\vecQ/\partial\vecmu \equiv \partial P/\partial \mu$ 
can be either positive or nonpositive (i.e., there are {\em two} possibilities), in the $s=2$ there are
{\em three} possibilities: \ when $\partial\vecQ/\partial\vecmu$ has two, one, or no 
positive eigenvalues. Thus, below we will give a geometric interpretation of these three
situations. Interestingly, this interpretation makes reference of a single curve (see \eqs~\eqref{e4_05}
below) --- an intersection line of surfaces $Q^{(1)}\big( \mu^{(1)}, \mu^{(2)} \big)$ and 
$Q^{(2)}\big( \mu^{(1)}, \mu^{(2)} \big)$ shifted in a certain manner.

For brevity, let us denote
\be
 \frac{\partial\vecQ}{\partial\vecmu} \equiv \left( 
  \ba{cc}  \partial Q^{(1)}/\partial \mu^{(1)} & \partial Q^{(2)}/\partial \mu^{(1)} \\
      \partial Q^{(1)}/\partial \mu^{(2)} & \partial Q^{(2)}/\partial \mu^{(2)}  \ea \right)
  \,=\, \left( \ba{cc} a_{11} & a_{12} \\ a_{12} & a_{22}  \ea \right)\,.
\label{e4_01}
\ee
Note that $\partial\vecQ/\partial\vecmu=R(0)$ is a symmetric matrix (see \eqref{e3_07}),
and so $a_{21}=a_{12}$. From the quadratic equation satisfied by its eigenvalues one can see that:
\bsube
\be
{\rm det}( \partial\vecQ/\partial\vecmu ) > 0 \;\; {\rm and} \;\; a_{11}, a_{22}>0 
  \quad \so \quad  p( \partial\vecQ/\partial\vecmu ) = 2;
\label{e4_02a}
\ee
\be
{\rm det}( \partial\vecQ/\partial\vecmu ) < 0 \hspace{3cm}
  \quad \so \quad  p( \partial\vecQ/\partial\vecmu ) = 1;
\label{e4_02b}
\ee
\be
{\rm det}( \partial\vecQ/\partial\vecmu ) > 0 \;\; {\rm and} \;\; a_{11}, a_{22}<0 
  \quad \so \quad  p( \partial\vecQ/\partial\vecmu ) = 0.
\label{e4_02c} \\
\ee
\label{e4_02}
\esube
We have used the symmetry of $\partial\vecQ/\partial\vecmu$ to infer that a definite sign of
$(a_{11}+a_{22})$ in (\ref{e4_02}a,c) implies the corresponding sign for {\em both} these
diagonal entries individually.

Let us consider two surfaces $Q^{(1)}\big( \mu^{(1)}, \mu^{(2)} \big)$ and 
$Q^{(2)}\big( \mu^{(1)}, \mu^{(2)} \big)$ and the normal vectors to them at a given point
$\big( \mu^{(1)}, \mu^{(2)} \big)$:
\be
\underline{n}_{\,k}=[a_{k1},\, a_{k2},\, -1]\,, \qquad k=1,2,
\label{e4_03}
\ee
where these vectors are chosen to point downward. If these surfaces are vertically shifted so
as to have the same height at point $\big( \mu^{(1)}, \mu^{(2)} \big)$, then the cross-product of
the normal vectors defines the direction of the intersection line of such shifted surfaces at this point. 
The vertical component of this cross-product is $ {\rm det}( \partial\vecQ/\partial\vecmu )$:
\be
\underline{n}_{\,1} \times \underline{n}_{\,2} = 
\left| \ba{ccr} \underline{i} & \underline{j} & \underline{k} \\
                a_{11} & a_{12} & -1  \\ a_{21} & a_{22} & -1 \ea \right| \,.
\label{e4_04}
\ee
Thus, according to the above definition, the intersection line, $\ell$,
of the shifted surfaces
$Q^{(1)}\big( \mu^{(1)}, \mu^{(2)} \big)$ and
$Q^{(2)}\big( \mu^{(1)}, \mu^{(2)} \big)$
points in the same vertical direction (i.e., up or down) as $\underline{n}_{\,1} \times \underline{n}_{\,2}$.
Let us also note that $(a_{k1},a_{k2})$ are the projections of $\underline{n}_{\,k}$ onto
the axes $\mu^{(1)}$ and $\mu^{(2)}$. With these observations, conditions \eqref{e4_02} are restated as:
\bsube
\be
\ba{l} \mbox{$\ell$ points up \ and} \\
      \mbox{projection of $\underline{n}_{\,k}$ on respective axis $\mu^{(k)}$ is {\em positive}} \ea
      \quad \so \quad \; p( \partial\vecQ/\partial\vecmu ) = 2;
\label{e4_05a}
\ee
\be
\mbox{$\ell$ points down}  \hspace{6.2cm} 
      \quad \so \quad p( \partial\vecQ/\partial\vecmu ) = 1;
\label{e4_05b}
\ee
\be
\ba{l} \mbox{$\ell$ points up \ and} \\
      \mbox{projection of $\underline{n}_{\,k}$ on respective axis $\mu^{(k)}$ is {\em negative}} \ea
      \quad \so \quad \; p( \partial\vecQ/\partial\vecmu ) = 0.
\label{e4_05c}
\ee
\label{e4_05}
\esube

Finally, let us note that conditions \eqref{e4_05} can be restated solely in terms of the
two-component vectors $\vec{a}_k \equiv [a_{k1},a_{k2}]$ obtained by projection of $\underline{n}_{\,k}$
on the horizontal plane:
\bsube
\be
\ba{l} \mbox{angle between $\vec{a}_1$ and $\vec{a}_2$ is \ $<\, 180^{\circ}$ \ and} \\
      \mbox{projection of $\vec{a}_k$ on respective axis $\mu^{(k)}$ is {\em positive}} \ea
      \quad \so \quad  p( \partial\vecQ/\partial\vecmu ) = 2;
\label{e4_06a}
\ee
\be
\mbox{angle between $\vec{a}_1$ and $\vec{a}_2$ is \ $>\, 180^{\circ}$}  \hspace{2.5cm} 
      \quad \so \quad \!\! p( \partial\vecQ/\partial\vecmu ) = 1;
\label{e4_06b}
\ee
\be
\ba{l} \mbox{angle between $\vec{a}_1$ and $\vec{a}_2$ is \ $<\, 180^{\circ}$ \ and} \\
      \mbox{projection of $\vec{a}_k$ on respective axis $\mu^{(k)}$ is  {\em negative}} \ea
      \quad \so \quad  p( \partial\vecQ/\partial\vecmu ) = 0,
\label{e4_06c}
\ee
\label{e4_06}
\esube
where the angle is measured from $\vec{a}_1$ to $\vec{a}_2$ in the counterclockwise direction.

\setcounter{equation}{0}
\section{Connection between convergence and dynamical stability}

First, we observe that the conditions \eqref{e3_01} under which the ITEM \eqref{e2_09} and
the modified CGM \eqref{A1_01} are guaranteed to converge \cite{footnote4}
are the same under which
the solitary wave of the incoherently coupled NLS-type equations \eqref{e3_18} 
with all nodeless components is linearly stable \cite{Pelinovsky2000}.
(More precisely, the system analyzed in \cite{Pelinovsky2000}
had $G^{(k)}=\sum_{l=1}^S \sigma_{kl}|U^{(l)}|^2$,
but the results of that paper are straightforwardly extended to apply to \eqref{e3_18}.)
Thus, the nodeless solutions of \eqref{e3_18} found by the iterative methods of this paper
are guaranteed to be dynamically linearly stable. This is
an extension of the result found in \cite{ITEM} for a single-component \eq~\eqref{e1_02}.

Let us now explain why this close connection between the convergence and stability takes place.
Our explanation applies both to single- and multi-component equations. In regards to
the convergence conditions, recall that the 
iteration methods converge when the operator $\Lcal$ in \eqref{e2_10} is negative definite
on the space of functions $\psi$ satisfying the orthogonality relation \eqref{e2_11}. Note that
on this space, \ $\langle \psi, \Lcal \psi \rangle = \langle \psi, {\bf L} \psi \rangle$, \
and therefore, the negative definitenesses of $\Lcal$ and ${\bf L}$ are equivalent 
under \eqref{e2_11}. 

Let us now turn to the stability conditions. The details are slightly different for envelope
and carrier solitary waves, so we begin with the former using \eqs~\eqref{e2_01} as an
example whenever needed. Seeking the slightly perturbed solitary wave in the form similar
to \eqref{e2_04} where now all $u^{(k)}$ are replaced by 
\be
u^{(k)} + \big(\tu^{(k)}_R + i \tu^{(k)}_I \big)e^{\lambda t}, \qquad u^{(k)}_{R,I}(\vecx)\in\mathbb{R},
\label{e5_01}
\ee
one obtains (see, e.g., \cite{Pelinovsky2000}):
\be
{\bf L} {\bf \tu}_R = \lambda {\bf \tu}_I, \qquad {\bf L}_I {\bf \tu}_I = -\lambda {\bf \tu}_R\,.
\label{e5_02}
\ee
In the case of \eqs~\eqref{e2_04} or \eqref{e3_18}, \ ${\bf L}_I={\bf \Lnot}$ (see \eqref{e2_08a}), 
but in general (e.g., for the three-wave system \cite{3w_76, 3w_02}) this is not so. 
Also, in the case where $s<S$, it is convenient, although not critical,
to write the $\vecmu$-term in ${\bf L}_I {\bf \tu}_I$ as, e.g., for \eqref{e2_05}: 
\ diag$\big( \mu^{(1)}, \mu^{(2)}, \mu^{(1)}, \mu^{(2)} \big)\,{\bf \tu}_I $; this makes
${\bf L}_I$ explicitly self-adjoint. 
What is important is that 
\be
{\bf L}_I \Ucal = {\mathcal O},
\label{e5_03}
\ee
where the right-hand side is the $S\times s$ zero matrix. Condition \eqref{e5_03} is, in fact, the
solvability condition of the second equation in \eqref{e5_02}, and is equivalent to the condition
that ${\bf \tu}_R$ satisfy the orthogonality relation \eqref{e2_11}:
\be
\langle \Ucal, {\bf \tu}_R \rangle = \vec{0}\,.
\label{e5_04}
\ee
Property \eqref{e5_04} is easily verified  by substituting \eqref{e5_01} into the conservation
law \ $d\vecQ/dt=\vec{0}$.

By \eqref{e5_04}, \eqref{e5_03} one can invert the second equation in \eqref{e5_02} and substitute
the result in the first equation:
\be
{\bf L} {\bf \tu}_R = -\lambda^2 {\bf L}_I^{-1} {\bf \tu}_R\,.
\label{e5_05}
\ee
Recall that this generalized eigenvalue problem is considered on the restricted space \eqref{e5_04}.
Both operators in \eqref{e5_05} are self-adjoint. Then, if ${\bf L}_I^{-1}$ (and hence
${\bf L}_I$) is negative definite, then by Sylvester's law of inertia,  the sign of $\lambda^2$
is determined by the signs of the eigenvalues of ${\bf L}$. If {\bf L} is negative definite
(again --- on the restricted space \eqref{e5_04}, or equivalently, \eqref{e2_11}), then $\lambda^2<0$
and hence the solitary wave is dynamically linearly stable (see \eqref{e5_01}).

Thus, to summarize: The conditions of convergence of the ITEM and modified CGM coincide with the
conditions under which the solitary wave is dynamically linearly stable if and only if operator
${\bf L}_I$ is negative definite. In particular, this occurs when ${\bf u}$ is a ground state. 
(Unfortunately,
the latter fact is not readily determined by inspection for multi-component solitary waves.)
Let us note that this result about the coincidence of convergence and stability conditions
for ground-state solitary waves 
fully agrees with the results of \cite{BaoD03, Bao04}, where it was proven that an ITEM-like
method converges to ground states of \eqref{e1_02} and its generalization \eqref{e3_18} describing
dynamics of Bose--Einstein condensates.

Finally, we give a counterpart of the above statement for carrier wave equations using
the KdV equation 
\be
u_t+2uu_x+u_{xxx}=0
\label{e5_06}
\ee
as an example. Its solitary wave $u=u(x-ct)\equiv u(\xi)$ with velocity $c$ satisfies an
equation  \ $u^2-cu+u_{xx}=0$, whose linearized operator is $L=2u-c+\partial^2_x$. The 
iterative methods seeking a solitary wave with a prescribed value of power \eqref{e1_01}
will converge provided that $L$ is negative definite on a space of functions $\psi$ satisfying a
variant of \eqref{e2_11}:
\be
\langle u, \psi \rangle =0.
\label{e5_07}
\ee
On the other hand, the stability analysis of \eqref{e5_06} via an ansatz 
\ $u(x,t)=u(\xi)+\tu(\xi)e^{\lambda t}$ \ leads to the eigenvalue problem
\ $\partial_x L\tu=\lambda\tu$. Upon the substitution $\tilde{w}=\partial_x^{-1}\tu$,
this eigenvalue problem is rewritten in the same form as \eqref{e5_02} \cite{KodPel05}:
\be
L\tu = \lambda \tilde{w}, \quad \L \tilde{w} = -\lambda\tu, \qquad \L\equiv -\partial_x L \partial_x\,.
\label{e5_08}
\ee
Note that due to the translational invariance of the solitary wave, $u$ is a solution of $\L u=0$.
Thus, all considerations for the envelope equations carry over to the case of \eqref{e5_06},
and we conclude that the convergence conditions of the ITEM and CGM coincide with the stability
conditions of the solitary wave if $u$ is the ground state of $\L$ (or, equivalently, $\L$ is
negative definite on the space defined by \eqref{e5_07}).

It may also be pointed out that in coupled multi-component generalizations of the KdV, 
there is only one parameter --- the wave's velocity $c$ --- that is the analog of the propagation
constant vector $\vecmu$ for the envelope equations, like in \eqref{e2_01} or \eqref{e3_18}.
Therefore, in this case, there is only one quadratic
conserved quantity (which is probably the sum of the
powers of all the components). In other words, $s=1$, and hence the ITEM and modified CGM can converge
only if $p({\bf L}) \le 1$.


\section{Conclusions}

In this work, we obtained the convergence conditions of the iterative methods \eqref{e2_09}
and \eqref{A1_01} that find multi-component solitary waves with prescribed values of
quadratic conserved quantities \eqref{e2_06}. These convergence conditions are given by
\eqref{e3_01} (provided that \eqref{e1_10} holds for the ITEM \eqref{e2_09}), which extend the 
convergence conditions of the single-component ITEM obtained in \cite{ITEM}. These conditions
also turn out to be the same as the dynamical
stability conditions for the nodeless (e.g., ground-state)
solitary wave of the system of incoherently coupled
NLS-type equations \eqref{e3_18}, which were obtained in \cite{Pelinovsky2000}. 
For a single-component NLS-type equation \eqref{e1_02}, such a coincidence was 
observed in \cite{ITEM}. Earlier, similar statements were proven for \eqref{e1_02} and
\eqref{e3_18} in \cite{BaoD03, Bao04} by different techniques.
In Section 5 we showed that, in general, 
the convergence conditions of the iterative methods \eqref{e2_09} and
\eqref{A1_01}, on one hand, and the dynamical stability conditions, on the other, coincide
for ground-state solitary waves of all Hamiltonian nonlinear wave equations.

Let us conclude with two remarks. First, even though we stated the ITEM \eqref{e2_09} 
in the main text of the paper while stating the CGM \eqref{A1_01} in Appendix, 
we remind the reader (see
Section 2) that if the ITEM converges slowly, then the modified CGM  will provide 
considerable acceleration of the iterations. Alternatively, one can use the slowest mode
elimination technique \cite{ME} to accelerate the algorithm of the ITEM. Comparison of these
three methods was done in \cite{CGM}, with the modified CGM being found the fastest.

Second, above we have explicitly mentioned the form of the operators employed by the
ITEM and modified CGM for envelope equations (like \eqref{e2_01} and
\eqref{e3_18}) and for the carrier waves (like the KdV, \eqref{e5_06}). 
For systems that couple envelope and carrier waves, 
which are commonly referred to as short--long wave interaction, or generalized
Zakharov--Benney, equations, the formalism remains the same.


\setcounter{equation}{0}
\renewcommand{\theequation}{A.\arabic{equation}}
\section*{Appendix: \ Modified CGM for solitary waves with a prescribed $\vecQ$}

The steps of this algorithm for \eq~\eqref{e2_08a} are (operator $\Lcal$ is defined in \eqref{e2_10}):
\bsube
\be
\vecr_0= \vecN^{-1} {\bf \Lnot  u}_0, \qquad 
\vecd_0=\vecr_0 -  \Ucal_0\, \langle \Ucal_0,\, \Ucal_0 \rangle^{-1}\, 
   \langle  \Ucal_0,\, \vecr_0 \rangle\,,
\label{A1_01a}
\ee
\be
\alpha_n=-
\frac{ \langle \vecr_n,\,\vecN \vecd_n \rangle }
{ \langle \vecd_n,\,\Lcal \vecd_n \rangle }\,,
\label{A1_01b}
\ee
\be
\hat{{\bf u}}_{n+1}={\bf u}_n+\alpha_n \vecd_n\,, \qquad
u^{(k)}_{n+1}=\hat{u}^{(k)}_{n+1} 
\sqrt{ \frac{ Q^{(k)}- \sum_{l=k+1}^S q^{(kl)} \hat{P}^{(l)}_{n+1} }{ \hat{P}^{(k)}_{n+1} } }\;, 
\qquad    k=1,\ldots\,,\, s\le S\,,
\label{A1_01c}
\ee
\be
\vecr_{n+1}= \vecN^{-1} {\bf \Lnot  u}_{n+1} \vspace{0.2cm}
\label{A1_01d}
\ee
\be
\beta_n= -
 \frac{ \dst \langle \vecr_{n+1},\,\Lcal \vecd_n \rangle - 
   \langle \Lcal \vecd_n,\, \Ucal_{n+1} \rangle 
   \langle \Ucal_{n+1},\, \Ucal_{n+1} \rangle^{-1}\,
    \langle \Ucal_{n+1}, \, \vecr_{n+1} \rangle  }
{ \langle \vecd_n,\,\Lcal \vecd_n \rangle },  \vspace{0.2cm}
\label{A1_01e}
\ee
\be
\vecd_{n+1}= \vecr_{n+1}+\beta_n \vecd_n - 
 \Ucal_{n+1} \, \langle \Ucal_{n+1},\, \Ucal_{n+1} \rangle^{-1}\, 
 \langle \Ucal_{n+1},\, \vecr_{n+1}+\beta_n \vecd_n \rangle  \,.
\label{A1_01f}
\ee
\label{A1_01}
\esube
Equation \eqref{A1_01a} defines the initial residual $\vecr_0$ and the search direction $\vecd_0$.
The first equation in \eqref{A1_01c} updates the iterative solution along the search direction $\vecd_n$
by making
a ``step" of ``length" $\alpha_n$ found in \eqref{A1_01b}. Equations (\ref{A1_01}d,f) update the
residual and the search direction using an auxiliary parameter $\beta_n$ computed in \eqref{A1_01e}.








%
%


\begin{thebibliography}{99}
\bibitem{footnote1}
In the language of Linear Algebra, $\Lcal$ differs from $L$ by a rank-one correction
(i.e., the last term in \eqref{e1_08}), and the relation between the eigenvalues of two
such matrices can be found, e.g., in \cite{TrefethenBau_NLA}, p. 230.
\bibitem{footnote2}
In \cite{ITEM} 
it was also required that for $p(L)=1$, the eigenfunction of $N^{-1}L$ 
corresponding to the positive eigenvalue be not orthogonal to $u$.
However, one can show, similarly to how we will do it in Section 3, that this special case 
actually falls under case \eqref{e1_12b}.
\bibitem{footnote4}
Recall that we always imply that \eqref{e1_10} holds for the ITEM.
\end{thebibliography}
\end{document}